\documentclass[aps,prl,twocolumn,showpacs,groupedaddress]{revtex4}
\usepackage{graphicx}
\usepackage{dcolumn}
\usepackage{bm}
\usepackage{amssymb}
\usepackage{amsmath}
\usepackage{amsfonts}

\begin{document}

\title{Friedel oscillations, impurity scattering and temperature dependence
of resistivity in graphene}
\author{ Vadim~V.~Cheianov and~Vladimir~I. Fal'ko }
\affiliation{\centerline{Physics Department, Lancaster University, Lancaster, LA1 4YB, UK}
}
\date{\today}

\begin{abstract}
\noindent We show that Friedel oscillations (FO) in grapehene are strongly
affected by the chirality of electrons in this material. In particular, the
FO of the charge density around an impurity show a faster, $\delta \rho \sim
r^{-3}$, decay than in conventional 2D electron systems and do not
contribute to a linear temperature-dependent correction to the resistivity.
In contrast, the FO of the exchange field which surrounds atomically sharp
defects breaking the hexagonal symmetry of the honeycomb lattice lead to a
negative linear T-dependence of the resistivity.
\end{abstract}

\maketitle

Screening strongly influences properties of impurities in metals
and semiconductors. While Thomas-Fermi screening suppresses the
long-range tail of a charged impurity potential, Friedel
oscillations (FO) of the electron density around a defect
\cite{Friedel} are felt by scattered electrons at a distance much
longer than the Thomas-Fermi screening length. Friedel
oscillations originate from the singular behavior
of the response function of the Fermi liquid at wave vector $2k_{\mathrm{F}}$%
. At zero temperature the decay of the amplitude $\delta \rho $ of these
oscillations with distance $r$ from the impurity obeys a power law
dependence. In a non-relativistic degenerate two-dimensional (2D)\ Fermi gas
\cite{Lau}, $\delta \rho \propto \cos (2k_{\mathrm{F}}r+\delta )/r^{2}$. In
2D electron systems Bragg scattering off the potential created by these
long-range FO strongly renormalises the momentum relaxation rate, $\tau
^{-1} $ for quasi-particles near the Fermi level, $\epsilon \approx \epsilon
_{\mathrm{F}}$, which leads to a linear temperature dependence of the
resistivity \cite{ZAN,DasSarma,GlazmanAleiner,Stern} in a 'ballistic'
temperature range $\epsilon _{\mathrm{F}}>T>h/\tau $ confirmed in recent
experiments on semiconductor heterostructures and Si field-effect
transistors \cite{Exp}.

Graphene-based transistor \cite{novo04,zhang05} is a recent
invention which attracts a lot of attention. Improvement of the
performance of this device requires identification of the dominant
source of electron scattering limiting its mobility. Those can be
structural defects of graphene lattice (vacancies and
dislocations), substitutional disorder, chemical deposits on
graphene and, importantly, charges trapped in or on the surface of
the underlying substrate.

In this Letter we investigate the effects of screening of scatterers in
graphene, and show how the phenomenon of FO can be used to gain insight into
the microscopic nature of disorder. Graphene (a monolayer of graphite) is a
gapless 2D semiconductor with a Dirac-like dispersion of carriers \cite%
{DresselhausBook,Ando}. In this material, it is the Thomas-Fermi screening
which is responsible \cite{NomuraMacDonald} for the experimentally observed
linear dependence of graphene conductivity on the carrier density \cite%
{novo04,zhang05}. Moreover, quasiparticles in graphene possess chiral
properties related to the sublattice composition of the electron wave on a
2D honeycomb lattice \cite{DresselhausBook}. Below we show that, due to the
latter peculiarity of graphene, FO of the electron density decay as $\delta
\rho \sim r^{-3}$ at a long distance from an impurity [faster than in a
usual 2D metal] and the linear temperature-dependent correction to the
resistivity,

\begin{equation}
R(T)-R(0)=-\frac{h}{e^{2}}\frac{T\hbar }{\epsilon _{\mathrm{F}}^{2}\tau
_{\ast }},  \label{R(T)}
\end{equation}%
is caused only by the FO of the exchange field and is strongly sensitive to
the microscopic origin of scatters. In\ Eq. (\ref{R(T)}), $\tau _{\ast }^{-1}
$ is the backscattering rate specifically from atomically sharp defects
distorting the hexagonal symmetry of the honeycomb lattice: structural
defects, chemical deposits and substitutional disorder. Note that Coulomb
scatterers do not contribute towards this result. Thus, we predict that in
graphene-based devices where scattering is dominated by charges trapped in
the substrate (or on its surface) the temperature dependent resitivity, $%
[R(T)-R(0)]/R\sim (T/\epsilon _{\mathrm{F}})(\tau /\tau _{\ast })$ is much
weaker than in conventional 2D semiconductor structures. We propose to use
the measurement of the resistivity in a temperature range $\epsilon _{%
\mathrm{F}}>T>h/\tau $ as a probe of microscopic composition of disorder.

\textit{Electrons in graphene} can be described using 4-component Bloch
functions [$\phi _{\mathbf{K}_{+},A}$, $\phi _{\mathbf{K}_{+},B}$, $\phi _{%
\mathbf{K}_{-,}B}$, $\phi _{\mathbf{K}_{-},A}$], which characterise the
electronic amplitudes on the two sublattices ($A$ and $B$) and valleys $%
\mathbf{K}_{+}$ and $\mathbf{K}_{-}$. Since for the "ballistic"
temperature regime $T\tau /h>1$ interference between waves
scattered from FO around different impurities can be neglected, we
study screening of an individual impurity. The single-particle
Hamiltonian in the presence of one impurity is
\begin{equation}
\hat{H}_{\mathrm{sp}}=-i\hbar v\mathbf{\Sigma }{ \cdot
}\mathbf{\nabla }+\hat{u}\,\delta (\mathbf{r}).  \label{H0}
\end{equation}%
In this expression $\mathbf{\Sigma }=(\Sigma _{x},\Sigma _{y})$ is a
two-dimensional vector whose components are $4\times 4$ matrices $\Sigma
_{x}=\Pi _{z}\otimes \sigma _{x}$ and $\Sigma _{y}=\Pi _{z}\otimes \sigma
_{y}$. Together with $\Sigma _{z}=\Pi _{0}\otimes \sigma _{z}$ they form
\cite{McCannWL} the generator algebra of the unitary group SU$_{2}^{\Sigma }$%
. Here $\Pi _{x,y,z}$ and $\sigma _{x,y,z}$ are sets of three Pauli matrices
acting on the valley and sublattice indices, respectively, and $\Pi _{0}$ ($%
\sigma _{0}$) are $2\times 2$ unit matrices acting on the valley
(sublattice) spaces. Another set of three matrices, $\Lambda _{x}=\Pi
_{x}\otimes \sigma _{z}$, $\Lambda _{y}=\Pi _{y}\otimes \sigma _{z}$ and $%
\Lambda _{z}=\Pi _{z}\otimes \sigma _{0}$, such that $[\Lambda
_{l_{1}},\Lambda _{l_{2}}]=2i\varepsilon ^{l_{1}l_{2}l}\Lambda _{l}$,
satisfies $[\Sigma _{s},\Lambda _{l}]=0$. Therefore, they commute with the
first term in $\hat{H}_{\mathrm{sp}}$ and generate the unitary group SU$%
_{2}^{\Lambda }\equiv \{\mathrm{e}^{ib\vec{n}\cdot \!\vec{\Lambda}}\}$
describing the valley symmetry of the Dirac Hamiltonian in graphene.

For a non-magnetic defect the matrix $\hat{u}\,$\ in Eq. (\ref{H0}) should
be hermitian and time-reversal symmetric. It can be parametrised using 10
independent real parameters \cite{McCannWL,AleinerEfetov}. One can check
\cite{McCannWL,timereversal} that all the operators $\Sigma _{s}$ and $%
\Lambda _{l}$ change sign under the time-reversal transformation, so that
the 9 products $\Sigma _{s}\Lambda _{l}$ are $t\rightarrow -t$ invariant and
together with the unit matrix $\mathrm{\hat{I}}=\Pi _{0}\otimes \sigma _{0}$
can be used as a basis to represent a non-magnetic static disorder:%
\begin{gather}
\hat{u}=u\mathrm{\hat{I}}+\sum_{s,l=x,y,z}u_{sl}\Sigma _{s}\Lambda
_{l}\,\equiv u\mathrm{\hat{I}}+X_{z}\Sigma _{z}+\mathbf{X}\cdot \mathbf{%
\Sigma },  \label{u} \\
X_{s}=\sum_{l=x,y,z}u_{sl}\Lambda _{l}.  \notag
\end{gather}%
The term $u\mathrm{\hat{I}}$ in Eq. (\ref{u}) represents an electrostatic
potential averaged over the unit cell. We attribute this term to charged
impurities with a sheet density $n_{\mathrm{c}}$. Various atomically sharp
defects \cite{Ziegler} with the 2D density $n_{\mathrm{def}}$ can be
parametrized using nine independent real parameters $u_{sl}$ and
time-inversion-symmetric \cite{timereversal} matrices $\Sigma _{s}\Lambda
_{l}$. In particular, $u_{zz}$ describes different on-site energies on the $%
A $ and $B$ sublattices. Terms with $u_{xz}$ and $u_{yz}$ take into account
fluctuations of $A\leftrightarrows B$ hopping, whereas $u_{sx}$ and $u_{sy}$%
\ ($s=x,y,z$) generate inter-valley scattering (whose presence has been
revealed \cite{McCannWL} by the observation of weak localisation in graphene
\cite{GeimWL}).

\textit{Thomas-Fermi screening.} The momentum relaxation of chiral electrons
in graphene due to scattering off Coulomb impurities is determined by the
anisotropic differential cross-section,
\begin{equation*}
w(\theta )\sim u_{k_{\mathrm{F}}\sin (\theta /2)}^{2}\cos ^{2}(\theta /2).
\end{equation*}%
The factor $\cos ^{2}(\theta /2)$, where $\theta $ is the scattering angle,
reflects the absence of back-scattering of graphene electrons off the
electrostatic potential \cite{AndoNoBS,Cheianov}. In the Thomas-Fermi
approximation, the Fourier transform $u_{q}$ of the electrostatic potential
of a single charged impurity screened by 2D electrons \cite%
{AndoFowlerStern} is $u_{q}=2\pi (e^{2}/\chi )/(q+\kappa )$, \ where $\kappa
=8\pi \gamma e^{2}/\chi =4k_{\mathrm{F}}r_{s}$; $\gamma =k_{\mathrm{F}%
}/(2\pi \hbar v)$ is the density of states per spin and valley, $\chi $ is
the dielectric constant, and $r_{s}\equiv e^{2}/\chi \hbar v$ (for a
graphene sheet placed on Si substrate, $r_{s}\sim 1$).

In a structure with sheet density $n_{\mathrm{c}}$ of charged impurities,
the resistivity, $R=[2e^{2}\gamma v^{2}\tau ]^{-1}$ is determined by the
momentum relaxation rate, $\tau ^{-1}\sim n_{\mathrm{c}}\pi \gamma \hbar
^{-1}\langle (1-\cos \theta )w(\theta )\rangle _{\theta }$, and it is
inverse proportional to the carrier density $n_{e}$ \cite{NomuraMacDonald},%
\begin{equation}
R=\frac{h}{4e^{2}}\frac{n_{\mathrm{c}}}{n_{e}}\frac{4\pi \eta
(r_{s})r_{s}^{2}}{(1+4r_{s})^{2}},  \label{R}
\end{equation}%
where $\eta (r_{s})$ is a monotonous function such that $\eta
(0)=1$, $\eta (1)\approx 0.3$ and $\eta (\infty )=\frac{1}{4}$. It
has been established that a contribution from atomically sharp
disorder towards $R$\ depends on the Fermi energy only
logarithmically \cite{AleinerEfetov}. Therefore, the empirical
relation $n_{e}R=\mathrm{const}$ established in the experimentally
studied graphene structures with high carrier densities
\cite{novo04,zhang05} indicates that charges located in the
underlying substrate or on its surfaces are the main source of
scattering. In Eq. (\ref{u}) it is represented by
$u\mathrm{\hat{I}}$.

\textit{Analysis of Friedel oscillations.}\textbf{\ }Below, we analyse the
FO of the density matrix of 2D electrons in graphene surrounding various
types of scatterers, Eq. (\ref{u}). The plane wave eigenstates of the Dirac
Hamiltonian $-i\hbar v\mathbf{\Sigma }{ \cdot }\mathbf{\nabla }$ are $%
\psi _{\mathbf{k},\xi }(\mathbf{r})=e^{i\mathbf{k}\mathbf{r}}|\mathbf{k}\xi
\rangle $. Here, the spinor $|\mathbf{k}\xi \rangle $ has a definite valley
projection, $\Lambda _{z}|\mathbf{k}\xi \rangle =\xi |\mathbf{k}\xi \rangle $%
, $\xi =\pm 1$ and chirality: the projection $\mathbf{\Sigma }\cdot \mathbf{n%
}=\pm 1$ of the operator $\mathbf{\Sigma }$ on the direction of motion, $%
\mathbf{n}=\mathbf{k}/k.$ Below, we describe a chiral electron
with $\mathbf{\Sigma }\cdot \mathbf{n}=1$ using the polarization
matrix
\begin{equation}
\hat{s}_{\mathbf{n}}\equiv \sum_{\xi =\pm 1}|\mathbf{k}\xi \rangle \langle
\mathbf{k}\xi |={\textstyle\frac{1}{2}}({\mathbf{1}}+\mathbf{\Sigma }\cdot {\mathbf{n)}}%
,\;\;\hat{s}_{\mathbf{n}}^{2}=\hat{s}_{\mathbf{n}}.  \label{PolMatrix}
\end{equation}

Due to scattering off the impurity, the plane wave eigenstates of the
Hamiltonian acquire a correction $\delta \psi _{\mathbf{k},\xi }$, which in
the Born approximation is given by
\begin{equation}
\delta \psi _{\mathbf{k},\xi }(\mathbf{r})=\int \frac{d^{2}p}{(2\pi )^{2}}%
\frac{e^{i\mathbf{p}\mathbf{r}}}{E(k)-\hbar v\mathbf{\Sigma }\cdot \mathbf{p}%
+i0}\hat{u}|\mathbf{k}\xi \rangle .  \label{dpsi}
\end{equation}%
Here, $E(k)=\pm \hbar vk$, where the sign $+(-)$ corresponds to the
conduction(valence) band states. For graphene with the Fermi energy $%
\epsilon _{\mathrm{F}}$ positioned in the conduction band, and for $%
T<\epsilon _{\mathrm{F}}$, we shall disregard the valence band contribution.
Then, the correction to the electron density matrix induced by the impurity
is
\begin{equation}
\delta \hat{\rho}(\mathbf{r},\mathbf{r}^{\prime })=\int \frac{d^{2}k}{(2\pi
)^{2}}n_{F}(k)\delta \psi _{\mathbf{k},\xi }(\mathbf{r})\otimes \psi _{%
\mathbf{k},\xi }^{\dagger }(\mathbf{r}^{\prime })+\mathrm{h.c.}  \label{drho}
\end{equation}

The matrix $\delta \hat{\rho}(\mathbf{r},\mathbf{r}^{\prime })$ contains a
slowly decaying oscillatory part, which is due to a jump in the Fermi
function $n_{\mathrm{F}}(k)$. After substituting Eq.~(\ref%
{dpsi}) into Eq.~(\ref{drho}) for the density matrix we find that
in the region $1<k_{\mathrm{F}}r<\epsilon _{\mathrm{F}}/T$, the
oscillating part of the 'local' density matrix,
$\hat{\rho}(\mathbf{r},\mathbf{r})$ calculated to leading order in
$1/r$ is given by
\begin{gather}
\delta \hat{\rho}(\mathbf{r},\mathbf{r})=\frac{k_{\mathrm{F}}}{8\pi ^{2}v}%
\frac{e^{2ik_{\mathrm{F}}r}}{2ir^{2}}\hat{s}_{\mathbf{n}}\hat{u}\hat{s}_{-%
\mathbf{n}}+\mathrm{h.c.};  \label{rhoans} \\
\hat{s}_{\mathbf{n}}\hat{u}\hat{s}_{-\mathbf{n}}={\textstyle \frac{1}{2}}(\hat{X}_{z}+i%
\mathbf{n}\times {\mathbf{\hat X)}}(\hat{\Sigma}_{z}-i\mathbf{n}\times \hat{%
\bm\Sigma }),  \notag
\end{gather}%
where $\mathbf{n}=\mathbf{r}/r$ and $\mathbf{a}\times
\mathbf{b}\equiv a_{x}b_{y}-a_{y}b_{x}$. The matrix structure of
$\delta \hat \rho$ describes the distribution of charge at the
atomic length scale: oscillations with the wave vector $\mathbf K
=\mathbf K_+- \mathbf K_-$ (related to $\Lambda_{x,y}$) superimposed
with the oscillations between the $A$ and $B$ sublattices (related
to, e.g., $\Sigma_z$).

The FO in the density matrix (\ref{rhoans}) do not lead \cite%
{Lin} to oscillations in the charge density, $\delta n_{e}(\mathbf{r})=%
\mathrm{Tr}\delta \hat{\rho}(\mathbf{r},\mathbf{r})$ because for any pair of
matrices $\Sigma _{s}$ and $\Lambda _{l}$, $\mathrm{Tr}\Sigma _{s}\Lambda
_{l}=0$. To find\ $\delta n_{e}$ we evaluated $\delta \hat{\rho}$ up to
order $r^{-3}$ in the $1/(k_{\mathrm{F}}r)$ expansion and arrived at the
leading non-vanishing contribution to the FO of the electron density coming
from the diagonal disorder, $u\mathrm{\hat{I}}$%
\begin{equation}
\frac{\delta n_{e}(\mathbf{r})}{n_{e}}=\frac{un_{e}}{\epsilon _{\mathrm{F}}}%
\frac{\cos (2k_{\mathrm{F}}r)}{(2k_{\mathrm{F}}r)^{3}},\;k_{\mathrm{F}}r\gg
1,  \label{dn}
\end{equation}%
where for a Thomas-Fermi screened impurity charge ($un_{e}/\epsilon _{%
\mathrm{F}})\sim 2r_{s}/(1+4r_{s})\sim 1$. The FO in Eq. (\ref{dn}) decay
with the distance from\ the Coulomb impurity faster than in a conventional
2D electron gas (where FO obey the $1/r^{2}$ law). This is due to the
absence of backscattering of a chiral electron off the potential $u\mathrm{%
\hat{I}}$ conserving the sublattice state \cite{AndoNoBS}. As a result,
Bragg scattering off FO formed around the Coulomb impurity is suppressed and
does not lead to a linear T-dependence of the resistivity.

\textit{Interaction correction to resistivity.}\textbf{\ }Since
atomically sharp defects do not generate FO in the electron
density, they also do not induce an oscillating Hartree potential,
though they generate a non-local exchange field. Below we
investigate using the Born approximation how the exchange field
created by the FO in Eq. (\ref{drho}) renormalises impurity
scattering and leads to a linear $T$-dependence of the
resistivity. We write down
the Hartree-Fock potential created by the impurity as a sum $H_{\mathrm{H}%
}+H_{\mathrm{F}}$,
\begin{equation*}
\begin{array}{l}
\displaystyle H_{\mathrm{H}}=2\int d\mathbf{r}d\mathbf{r}^{\prime }V(\mathbf{%
r}-\mathbf{r}^{\prime })\mathrm{Tr}\rho (\mathbf{r}^{\prime },\mathbf{r}%
^{\prime })\psi ^{\dagger }(\mathbf{r})\psi (\mathbf{r}), \\
\displaystyle H_{\mathrm{F}}=-\int d\mathbf{r}d\mathbf{r}^{\prime }V(\mathbf{%
r}-\mathbf{r}^{\prime })\psi ^{\dagger }(\mathbf{r})\rho (\mathbf{r},\mathbf{%
r}^{\prime })\psi (\mathbf{r}^{\prime }),%
\end{array}%
\end{equation*}%
where $V(\mathbf{r})$ is the (screened) Coulomb potential averaged over the
unit cell \cite{footnote}. Here, we suppressed the electron spin indices,
for brevity. The factor of $2$ in $H_{\mathrm{H}}$ is the result of
summation over spin channels.

\begin{figure}[tbp]
\includegraphics
[width =0.35 \textwidth] {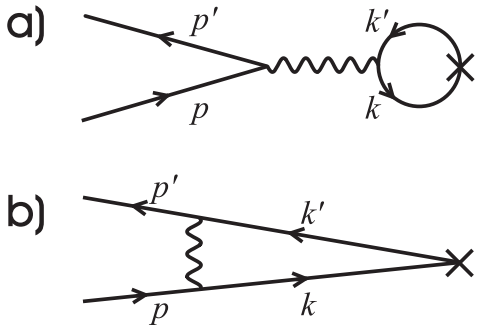} \caption{The figure is a
schematic representation of the processes renormalizing the
scattering amplitude in the Hartree a) and the Fock b) channel.}
\label{fig:diagrams}
\end{figure}

The leading correction $\delta w(p,\theta )$ to the differential
cross-section $w(\theta )$ of scattering off the screened defect is a result
of the interference \cite{ZAN} between the wave scattered by the defect
itself and the wave Bragg-reflected by the Hartree, Fig.~\ref{fig:diagrams}%
(a) and the Fock, Fig.~\ref{fig:diagrams}(b) potentials of the FO,
The resulting correction to the momentum relaxation rate is
\begin{equation}
\delta \tau ^{-1}=-n_{\mathrm{def}}v\int dpn_{\mathrm{F}}^{\prime }(p)\int
\frac{d\theta }{2\pi }(1-\cos \theta )\delta w(p,\theta ).  \notag
\end{equation}%
After substituting the Born amplitude, $A(\theta )$ of the scattering
process of the defect itself and the amplitudes, $\delta A_{a(b)}(\theta )$
of the two processes in Fig. 1 (a) and (b) into the differential
cross-section $w(\theta )=|A+\delta A_{a}+\delta A_{b}|^{2}$ one finds that $%
\delta w(\theta )=2\mathrm{\!}\Re \lbrack A^{\ast }(\delta A_{a}+\delta
A_{b})]$, which yields

\begin{widetext}
\begin{align}
& \delta \tau^{-1} = \frac{n_{\rm def} }{2\pi^4 v^2} \Re
\int d\mathbf p d \mathbf p' d \mathbf k d \mathbf k'
\delta_E\delta_M \frac{1-\mathbf{nn'}}{k^2-(k')^2+i 0}
\left[\tilde V(\vert \mathbf k-\mathbf p\vert) k \Gamma_{\mathbf
p, \mathbf p'}^{\mathbf k, \mathbf k'} - 2 \tilde V(\vert \mathbf
p'-\mathbf p\vert) k \tilde \Gamma_{\mathbf p, \mathbf
p'}^{\mathbf k, \mathbf k'} \right] n_{\rm F}(k) n_{\rm F}'(p),
\label{tauinv}
\\
\nonumber
& \text{where } \qquad
 k  \Gamma_{\mathbf p, \mathbf p'}^{\mathbf k, \mathbf k'} ={\textstyle
 \frac{1}{2}}
\mathrm{ Tr} \left[\hat s_{\mathbf n} \hat u \hat s_{\mathbf n'}
(k+ \bm \Sigma\cdot \mathbf k')\hat u \hat s_{\mathbf m}\right],
\qquad
 k \tilde \Gamma_{\mathbf p, \mathbf p'}^{\mathbf k, \mathbf k'}  ={\textstyle
 \frac{1}{2}}
 \mathrm{ Tr} \left[\hat s_{\mathbf n}
\hat u \hat s_{\mathbf n'}\right]\,
\mathrm{Tr} \left[(k+ \bm \Sigma\cdot \mathbf k')
\hat u \hat s_{\mathbf m}\right].
\end{align}
\end{widetext}

In Eq.~\eqref{tauinv}, the integration is constrained by the momentum and
energy conservation imposed by $\delta _{M}=\delta ^{2}(\mathbf{p}-\mathbf{p}%
^{\prime }+\mathbf{k}^{\prime }-\mathbf{k})$ and $\delta _{E}=\delta
(p-p^{\prime })$ and we use the notations $\mathbf{n}=\mathbf{p}/p$, $%
\mathbf{n}^{\prime }=\mathbf{p}^{\prime }/p$, $\mathbf{m}=\mathbf{k}/k.$ Two
form-factors of Bragg scattering, $\Gamma $ and $\tilde{\Gamma}$ represent
the Fock and the Hatree contributions, respectively. Assuming $x-y$ isotropy
of disorder, we average the relaxation rate $\tau ^{-1}$ over the directions
of the initial wave vector $\mathbf{p}$.

\begin{figure}[tbp]
\includegraphics
[width =0.35 \textwidth] {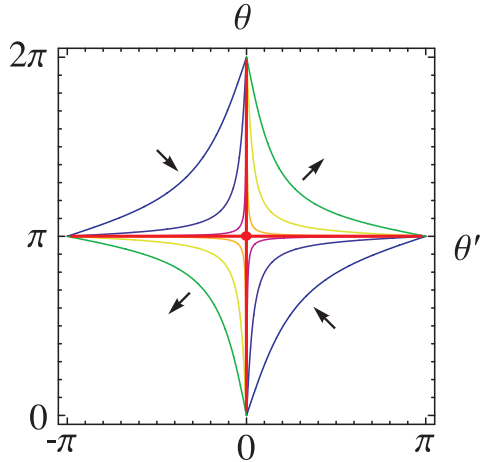} \caption{The structure of
singularities in the integration. The contours
show the position of the singularities of the integrand in the $\protect%
\theta ,\protect\theta ^{\prime }$ plane. The arrow indicate the direction
of increasing $k/p.$}
\label{fig:singularity}
\end{figure}

The Bragg scattering correction to the momentum relaxation rate %
\eqref{tauinv} can be separated into zero-temperature and
temperature-dependent parts, $\delta \tau ^{-1}=\delta \tau _{0}^{-1}+\delta
\tau _{T}^{-1}$. The linear temperature dependence of $\delta \tau _{T}^{-1}$
\cite{ZAN} is due to a singularity at $k=k^{\prime }$ in the integral %
\eqref{tauinv}. To evaluate its contribution we extend the analysis
presented in Ref. \cite{ZAN}. First, we use $\delta _{M}$ and $\delta _{E}$\
to rewrite the denominator in the integral \eqref{tauinv} as $%
k^{2}-(k^{\prime })^{2}=2(\mathbf{k}-\mathbf{p})(\mathbf{p}-\mathbf{p}%
^{\prime })=2kp(\cos \theta -\cos \theta ^{\prime })-2p^{2}[1-\cos (\theta
-\theta ^{\prime })]$, where $\theta =\widehat{\mathbf{kp}}$ and $\theta
^{\prime }=\widehat{\mathbf{kp}^{\prime }}$. The integrand is singular at
those points where the denominator vanishes, except for $\theta =\theta
^{\prime }$ where the singularity is cancelled by the factor $[1-\cos
(\theta -\theta ^{\prime })]$. For given $k$ and $p$ the locus of singular
points is a contour in the plane $(\theta ,\theta ^{\prime })$ which we show
in Fig.~\ref{fig:singularity} for various ratios $k/p$. As $k$ changes from $%
k<p$ to $k>p$, two parts of this contour coalesce creating a
double pole at the point $(\theta ,\theta ^{\prime })=(0,\pi )$.
It is convenient to rewrite Eq. \eqref{tauinv} as
\begin{equation*}
\delta \tau _{T}^{-1}=\frac{n_{\mathrm{def}}}{4\pi ^{4}v^{2}}\int dpdk\,n_{%
\mathrm{F}}(k)n_{\mathrm{F}}^{\prime }(p)g(k,p)-\delta \tau _{0}^{-1},
\end{equation*}%
where the function $g(k,p)$ is the result of the integration of Eq.~%
\eqref{tauinv} over $\mathbf{p}^{\prime }$ and $\mathbf{k}^{\prime }$ and
the angular components of $\mathbf{p}$ and $\mathbf{k}$. The leading term in
$\delta \tau _{T}^{-1}$ is linear in temperature because of the jump, $%
g(p+0,p)-g(p-0,p)=\Delta g(p)$ of the function $g$ at $k=p$ resulting from
the double pole at $(\theta ,\theta ^{\prime })=(0,\pi )$ for $k=p$:
\begin{gather}
\delta \tau _{T}^{-1}=-\frac{n_{\mathrm{def}}}{4\pi ^{4}v^{2}}\frac{\Delta
g(k_{\mathrm{F}})}{v}T=-T\frac{2n_{\mathrm{def}}k_{\mathrm{F}}}{\pi \hbar
^{4}v^{3}}\Gamma _{0}\tilde{V}(0),  \label{tauT} \\
\Gamma _{0}=\int \frac{d\mathbf{n}}{2\pi }\mathrm{\ Tr}\left[ \hat{u}\hat{s}%
_{\mathbf{n}}\hat{u}\hat{s}_{-\mathbf{n}}\right] =\!%
\sum_{s,l=x,y,z}u_{sl}^{2}(1+\delta _{sz}).  \notag
\end{gather}

When deriving $\Delta g(k_{\mathrm{F}})$ in Eq. (\ref{tauT}), we took into
account that the contribution to the jump of $g$ comes from backscattering, $%
(\theta ,\theta ^{\prime })=(0,\pi )$, so that we substituted $\mathbf{p}=%
\mathbf{k}$ and $\mathbf{p}^{\prime }=-\mathbf{p}$ in all smooth
angle-dependent functions. This produced the exchange term form-factor $%
\Gamma _{0}\equiv \left\langle \Gamma _{\mathbf{p},-\mathbf{p}}^{\mathbf{p},-%
\mathbf{p}}\right\rangle _{\mathbf{p}}$ averaged over the incidence angle,
and also resulted in the vanishing of the Hartree term due to the absence of
backscattering of chiral electrons off the Coulomb potential, $\tilde{\Gamma}%
_{\mathbf{p},-\mathbf{p}}^{\mathbf{p},-\mathbf{p}}=0$.

Finally, we arrive at the following expression for the dominant
temperature-dependent part of the resistivity,

\begin{equation}
R(T)-R(0)=-\frac{h}{e^{2}}\frac{4T\tilde{V}(0)}{hv^{2}\epsilon _{\mathrm{F}%
}\tau _{\ast }},~~\tau _{\ast }^{-1}=\frac{\pi \gamma n_{\mathrm{def}}}{%
\hbar }\Gamma _{0}.  \label{FinalResult}
\end{equation}%
It is interesting to compare the latter result with that derived \cite{ZAN}
for electrons in a simple-band 2D semiconductor or metal. In the latter
case, $R(T)$ is formed by the competition of the Hartree and Fock
contributions which have different signs, which may lead to the change of
the size and even the sign of the effect upon variation of the electron
density or spin polarisation of the 2D gas by an external magnetic field. In
graphene the $T$-dependent correction to the resistivity is only due to the
exchange interaction, so that it is negative and its density dependence
tracks the density dependence of the interaction constant $\tilde{V}(0)$.
For a 2D-screened Coulomb interaction, $\tilde{V}(q)=2\pi (e^{2}/\chi
)/(q+\kappa )$, we estimate $\tilde{V}(0)=\frac{1}{4}\gamma ^{-1}$, which
leads to the result in Eq. (\ref{R(T)}). The other difference between
graphene and usual semiconductor structures is that in the former the linear
in $T$ correction is caused only by atomically sharp disorder (\textit{e.g.}%
, structural defects in graphene, chemical deposits,
substitutional disorder and contact with an incommensurate lattice
of a substrate), whereas in the latter it is formed by all
scatterers. This means that the temperature dependence of
resistivity should be weak (or even absent) in a suspended
graphene sheet or a sheet loosely attached to the substrate, and
that it can be pronounced in devices where graphene is strongly
coupled to the substrate.

\textit{In conclusion}, we described the Friedel oscillations
induced by scatterers in graphene and their effect on resistivity.
The general form of FO is given in Eq.~\eqref{rhoans}. One of the
immediate implications of this result is a $r^{-2}$ dependence of
the RKKY interaction between two magnetic impurities in graphene.
This is because a substitution atom with spin $\mathbf S$ will
generally create a perturbation $(\mathbf S \cdot \mathbf s_{e})\hat
u$ for the electron spin $\mathbf s_e$ with $\hat u$ containing all
symmetry-breaking terms in Eq.~\eqref{u}. We also showed that due to
the chirality of graphene electrons, the linear temperature
correction to the resistivity [ Eqs. \eqref{R(T)} and
\eqref{FinalResult}] is caused by atomically sharp defects rather
than by Coulomb charges in the insulating substrate, so that its
measurement can be used as a tool to test the microscopic
composition of disorder.

We thank I.Aleiner and A.Ludwig for discussions. This project was
funded by the EPSRC grant EP/C511743.


\begin{thebibliography}{99}
\bibitem{timereversal} Corners of the hexagonal Brilloin zone are $\mathbf{K}%
_{\xi }=\xi ({\textstyle\frac{4}{3}}\pi a^{-1},0)$, where $\xi =\pm 1$ and $%
a $ is the lattice constant. In the basis\ [$\phi _{\mathbf{K}_{+},A}$, $%
\phi _{\mathbf{K}_{+},B}$, $\phi _{\mathbf{K}_{-,}B}$, $\phi _{\mathbf{K}%
_{-},A}$], time reversal, $T(W)$ of an operator $W$ is described by $T(\hat{W%
})=(\Pi _{x}\otimes \sigma _{x}){W}^{\ast }(\Pi _{x}\otimes \sigma _{x})$.
This can be used to show that $T(\Sigma _{s})=-\Sigma _{s}$, $T(\Lambda
_{l})=-\Lambda _{l}$, and $T(\Sigma _{s}\Lambda _{l})=\Sigma _{s}\Lambda
_{l} $.
\bibitem{footnote} We neglect the non-conservation of the sublattice and
valley indices in the e-e scattering. The corrections due to such processes
are small, as $V(2\pi /a)/V(k_{\mathrm{F}})\sim k_{\mathrm{F}}a\ll 1$, where
$a$ is the lattice constant.
\end{thebibliography}
\end{document}